# Time reversal of ultrasound in granular media

Maxime Harazi, Yougu Yang, Mathias Fink, Arnaud Tourin[1], and Xiaoping Jia[2]

Institut Langevin, ESPCI Paris, PSL Research University, CNRS, 75005 Paris, France

## Abstract

Time reversal (TR) focusing of ultrasound in granular packings is experimentally investigated. Pulsed elastic waves transmitted from a compressional or shear transducer source are measured by a TR mirror, reversed in time and back-propagated. We find that TR of ballistic coherent waves onto the source position is very robust regardless driving amplitude but provides poor spatial resolution. By contrast, the multiply scattered coda waves offer a finer TR focusing at small amplitude by a lens effect. However, at large amplitude, these TR focusing signals decrease significantly due to the vibration-induced rearrangement of the contact networks, leading to the breakdown of TR invariance. Our observations reveal that granular acoustics is in between particle motion and wave propagation in terms of sensitivity to perturbations. These laboratory experiments are supported by numerical simulations of elastic wave propagation in disordered 2D percolation networks of masses and springs, and should be helpful for source location problems in natural processes.

## 1. Introduction

In a non-dissipative medium, the wave equation is symmetric in time. Therefore, for every wave diverging from a pulsed source, there exists in theory a wave, the time-reversed wave, that precisely retraces all its original paths in a reverse order and converges in synchrony at the original source as if time were going backwards. This time-symmetry exists even in a strongly heterogeneous medium where waves are strongly reflected, refracted, or scattered. In the early nineties, an original method for generating such a time-reversed wave was proposed in acoustics [1]: a pulsed wave is sent from a source, propagates in an unknown media and is captured at a transducer array termed a "Time Reversal Mirror (TRM)". Then the waveforms received at each transducer are reversed in time and sent back, resulting in a wave converging at the original source regardless of the complexity of the propagation medium. TRMs have now been implemented in a variety of physical scenarios from hundreds of Hz in ocean acoustics [2] and MHz Ultrasonics [3] to GHz Microwaves [4]. Common to this broad range of scales is a remarkable robustness exemplified by observations that the more scattering the medium, the sharper the focus [2-7].

---

[1] arnaud.tourin@espci.fr
[2] xiaoping.jia@espci.fr

For the last decade the time reversal focusing concept has also been at the heart of very active research in seismology, especially for seismic source imaging and source location of seismic events that exhibit no compressional ($P$-) and shear ($S$-) wave arrivals, such as tremor, glacial earthquakes and Earth hum [8,9]. In that case the real Earth, i.e., the medium where the wave field is generated and propagates, and the virtual Earth, i.e., the velocity model in which the time-reversed wave is numerically back-propagated, are however different.

As a model system for athermal amorphous media or seismic fault gouges, the granular medium constitutes a particular case among strongly scattering systems [10-12]. Dry granular media are collections of macroscopic grains that interact through repulsive and frictional contact forces. For given values of macroscopic control parameters, such as packing density and confining pressure, granular media exhibit many microstates characterized by highly heterogeneous contact force networks that can rearrange under driving. These media whose features range from the microscopic scale (grain) to the mesoscopic scale (force-chain) and the macroscopic scale (bulk), may be modelled either as particulate or continuum materials [13].

Elastic waves that propagate from grain to grain provide a unique probe of the contact force networks. Generally speaking, one distinguishes between the long-wavelength coherent ($P$- and $S$-) waves and the short-wavelength waves scattered by the heterogeneous force chains [11], often referred to as coda waves. The study of the TR focusing of elastic waves in a granular medium raises two challenging issues. First, no wave equation is available on the scale of the force chains. This issue is related to a fundamental question: on what scale is the continuum elasticity applicable in a contact network [14-16]? Secondly, one may wonder whether time-reversal invariance still holds in a fragile granular medium, beyond a certain wave amplitude where the wave itself not only acts as a probe but also as a pump, leading to the acoustic fluidization of the jammed media via the rearrangement of the contact network [17,18]. This situation is fundamentally different from those previously reported where a perturbation was introduced in the continuous medium between the forward and backward propagation steps [6,7].

In this work, we address the above issues by experimentally investigating time-reversal focusing of ultrasonic waves in glass bead packings under external load. The robustness of TR invariance is tested with a specifically developed TRM as a function of the source amplitude. A particular attention is paid to the spatial extent of the rearrangement caused by large-amplitude driving.

## 2. Experiments

A sketch of the experimental setup is shown in Fig. 1a. Our granular materials consists of dry monodisperse glass beads of diameter $d$ = 1.5 or 3 mm, confined in a cylindrical container of diameter $D$ = 150 mm with rigid walls (i.e., an oedometer cell) which is filled to a height of $H$ = 55 mm with a packing density of about $\phi$ = 0.62. A static uniaxial stress $P \approx 85$ kPa is applied to the granular packing. To perform the time-reversal experiment, a compressional or shear transducer is placed in contact with the granular packing at the top of the cell and used as a source for transmitting a 3-cycle tone burst centered at $f$ = 100 kHz. We have developed a specific time-reversal mirror (TRM) with sixteen identical transducers, compressional or shear. Six other transducers surrounding the source (with a pitch of 20 mm between neighbouring transducers) are used to measure the extension of the time-reversed focal spot. The diameters of these transducers are 12 mm, which are much larger than the bead size $d$ to ensure an efficient detection of transmitted elastic waves. To investigate nonlinear effects, we vary the source amplitude from 5 to 300 V, corresponding to a vibration displacement $u_0 \approx 1 - 60$ nm [18].

*TR in the linear regime* –Fig. 1b depicts a typical waveform transmitted through a packing of 1.5-mm-diam glass beads, excited by a longitudinal transducer at small amplitude ($u_0$ < 5 nm) and measured by one of four transducers located at the centre of the TRM. We clearly identify the early arrival of a low-frequency coherent ballistic P wave ($f_{LF}$ ~ 15 kHz), followed by high-frequency waves ($f_{HF}$ ~ 80 kHz) resulting from the scattering by the heterogeneous force chains, i.e., coda waves. The compressional wave velocity can be measured by the time-of-flight of the P-wave pulse as $V_P \approx 500$ m/s, which gives a wavelength $\lambda$ of about 33 mm and 6 mm ($\sim 4d$) for the long-wavelength coherent and short-wavelength coda waves, respectively [11].

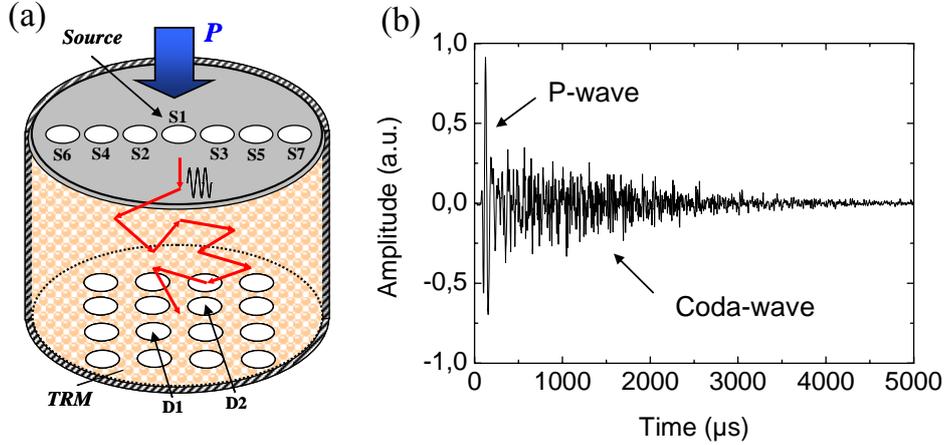

FIG. 1 (a) Experimental setup: glass beads are placed between the source S1 and the 16-element TRM. Detectors S2 to S7 allow for measurements of the spatial focusing. (b) A typical ultrasonic waveform transmitted through a packing of glass beads with a diameter of $d$ = 1.5 mm detected at bottom by D1. It consists of a low-frequency coherent wave followed by high-frequency scattered waves.

In a first TR experiment, only the coherent wave received by the TRM (16 transducers) is time-reversed and back-propagated. As shown in Fig. 2a, the time-reversed signal is focused at the source location around the focal time $t \approx 0$ (arbitrary) but with a spot much larger than the source size. In a second series of experiments, only the coda waves are selected, time-reversed and back-propagated. The TR focal spot is found to be finer with the scattered coda waves (Fig. 2b) than with the coherent wave (Fig. 2a), providing a higher resolution of the source location. In a homogeneous medium, diffraction theory predicts a focal spot of size around $\lambda H/D$ where $D$ is the aperture of the TRM and $H$ the focusing distance. To ascertain that the fine TR focusing provided by coda waves is not simply associated with its higher frequency content, we have performed the same TR focusing experiment with a single-transducer TRM. In that case no focusing is expected for the low-frequency coherent wave but we found that time-reversed coda waves are still focused. Such observation is in agreement with previous TR experiments performed in strongly multiple scattering media [3,7] or chaotic cavity [5]: multiple scattering allows for redirection of the source angular spectrum towards the TRM, which amounts to creating a virtual aperture larger than the actual TRM aperture –a kind of lens effect [3,5]. Notice that TR of the long-wavelength coherent wave is very robust to external perturbations, e.g., by a gentle tapping between the forward and the backward propagation steps, whereas the short-wavelength coda waves are not.

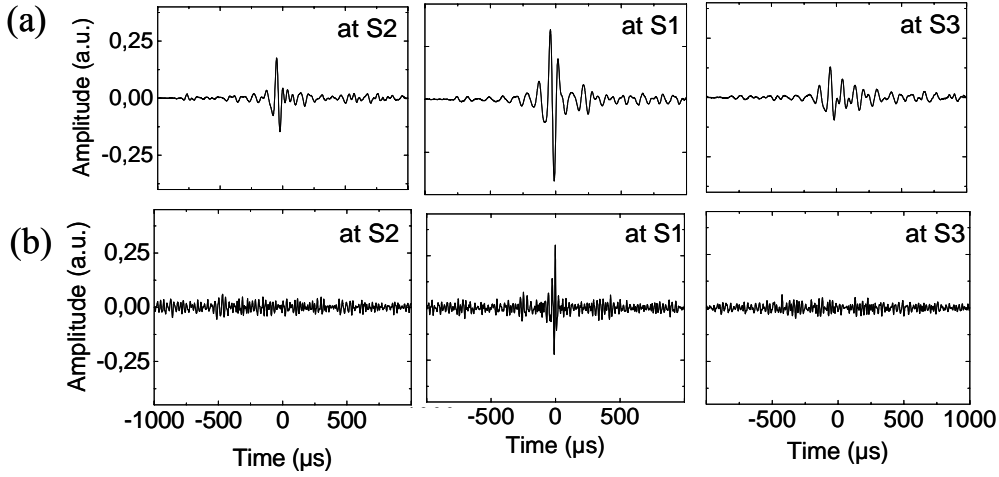

FIG. 2 Recompression signals recorded at the source location S1 and at the closest neighbouring detectors S2 and S3 after time-reversal and back-propagation of (a) the low-frequency coherent wave and (b) the high-frequency multiply scattered waves in bead packings ($d$ = 1.5 mm)

*TR in the nonlinear regime* –In the following, we concentrate on the source localization by the TR focusing of multiply scattered waves. To this end, we conduct TR experiments in packings of 3-mm-diam glass beads where the elastic wave transmission is dominated by scattered coda waves with $\lambda/d$ ~ 2 (Fig. 3a) with a transport mean free path $l^*$ ~ $\lambda$ [12]. To measure the fidelity of the wave reconstruction, we follow the pulse recompression, at the source location as a function of the driving amplitude for either a compressional or shear source. The quality of this reconstruction can be evaluated through the signal-to-noise ratio (*SNR*) defined as the peak amplitude of the TR recompression at the focal time divided by the standard deviation (RMS) of the symmetrically surrounding side-lobes calculated in a time-window of arbitrary length (rectangular boxes in Fig. 3b). Here it is important to point out that these side-lobe signals are not due to the electronic noises but to the imperfections of the TR focusing [3,5], and therefore the SNR is a clear indicator of the wave fidelity.

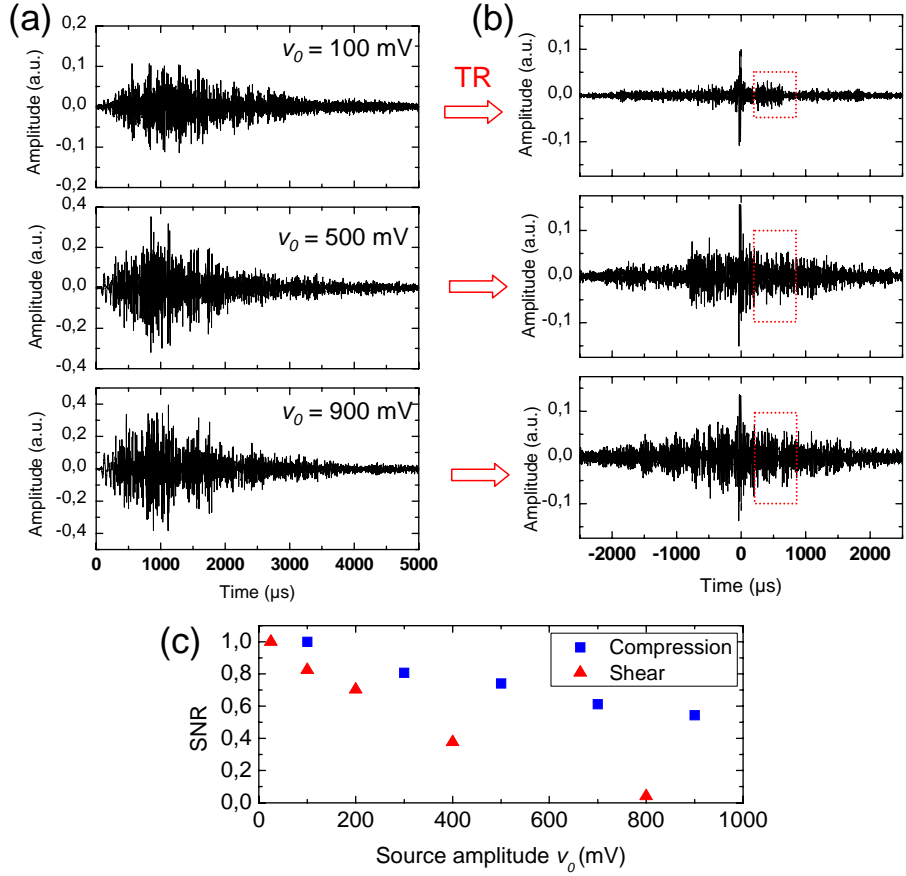

FIG 3. The (longitudinal) source amplitude is gradually increased in bead packings ($d$ = 3 mm). (a) Typical waveforms received at the TRM excited by various source amplitudes $v_0$ (corresponding to $u_0$ ~ 6, 30, 54 nm). (b) TR signals measured at the source positions where (red) boxes denote the range of side lobes for the RMS estimation. (c) Ratio of the TR signal at the focal time to RMS versus $v_0$ both for a compression and shear acoustic source.

We have that verified both the TR signals at the focal time and their side-lobes increase with the source amplitude $v_0$ in the linear regime when $v_0$ < 50 mV (or $u_0$ < 5 nm) (data not shown), giving rise to a constant *SNR* (taken as 1 in Fig. 3c). However, in the nonlinear regime ($u_0$ > 6 nm), the peak amplitude starts to increase in a slower way (and tends to saturation) than the side-lobes amplitude, resulting in a decrease of *SNR* with increasing the source amplitude. Such effect is even more pronounced with a shear source transducer as seen in Fig. 3c. The loss in the fidelity of the TR focusing process is likely associated with the breakdown of the TR invariance between the forward and backward propagation steps, caused by the rearrangement of the contact network (without visible grain motion) induced by the large source amplitude [18]. This scenario is also consistent with the difference observed between transmitted coda signals excited at small and large source amplitudes, respectively, during the forward propagation step (Fig. 3a).

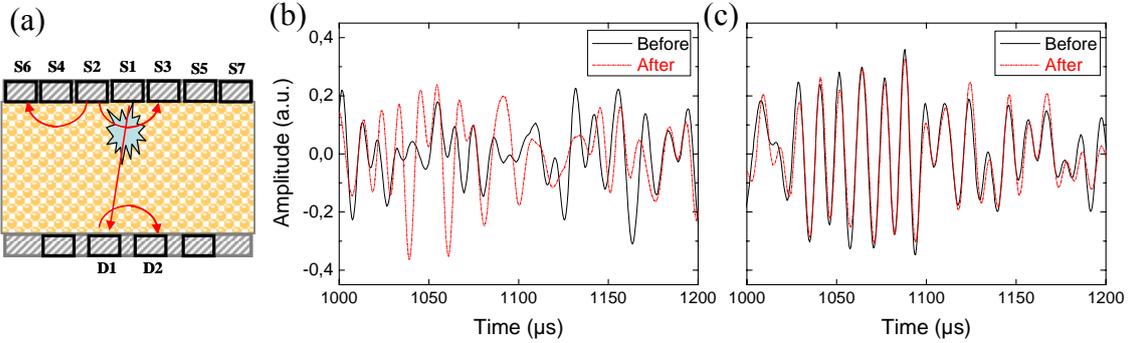

FIG. 4 Comparison of small-amplitude multiply scattered coda signals recorded before and after the large source driving. (a) These coda waves travel through different zones surrounding the source transducer. The degree of similarity $\Gamma$ inferred from correlation of coda waves shows that there are (b) the important vibration-induced structural changes around the source ($\Gamma \approx 0.42$) from S2 to S3, (c) but almost no modification away from the source ($\Gamma \approx 0.93$) from S2 to S6 (see text)

In order to evaluate the spatial extent of the network rearrangement caused by the strong source vibration, we examine the *local* change of the contact networks by using correlation of the configuration-sensitive multiply scattered waves, i.e., acoustic speckles [11,19]. More precisely, we measure the normalized correlation function $\Gamma$ that quantifies the degree of similarity of two successive, small-amplitude multiply scattered waves (used as nondestructive acoustic probes) recorded before and after the large-amplitude driving [18]; these waves are transmitted through the different zones of the granular packing surrounding the source S1 (Fig. 4). We observe that the multiply scattered waves crossing the source location zone, from transducers S1 to D1 or from S2 to S3, exhibit an important decorrelation $\Gamma < 0.4$ (Fig. 4b). On the other hand, the scattered waves travelling in zones away from the source, e.g. from S2 to S6 on the source side, or from D2 to D1 on the TRM side, remain highly correlated $\Gamma > 0.9$ (Fig. 4c). These results clearly indicate that the vibration-induced rearrangement of the contact networks in the nonlinear regime takes place primarily in the vicinity of the source and the fidelity loss of the TR focusing is mainly due to the structural change of granular packings between the forward and backward propagation steps.

## 3. Numerical simulations

The range of frequency used in this study lies well below the first resonances of a 3-mm-diam glass bead (the shear-like spheroidal mode arises at $f_{res} = (V_S)_{glass}/d \sim 1$ MHz). Thus the granular network can be modelled as an effective random network of point masses (beads) and springs [20], which exhibits spatial fluctuations of both density and elastic modulus. However, in the regime of multiple wave scattering, the elastic wave equations deduced from first principles calculation [21] are not available for such amorphous-like granular media. Various numerical simulations using molecular dynamics (MD) or discrete element method (DEM) based on the frictional Hertzian interaction have thus been used to model the wave propagation through granular packings [22,23].

*Description of the model* –Here, for simplicity, we investigate the TR focusing based on a toy model: a 2D percolation network of point masses connected by springs in which the structural disorder is obtained by randomly placing the masses $m$ on the simple square lattice sites with a fraction $p$ of the sites occupied. Using a uniform spring constant $k$, such a percolation network in 3D was previously used to simulate the heat diffusion (phonon transport) in amorphous solids where the amount of disorder can be controlled through $p$ (as the number density) [24]. To account for the heterogeneous network of the contact force (and stiffness) in a granular packing [13-16], we add disorder in the distribution of spring constants which are randomly chosen from a uniform distribution between 0 and $2k_0$. Fig. 5a and 5b depict sketches of a typical 2D percolation network where a given mass interacts with eight neighbours in general via different $k$; if we fill the lattice sites with monodisperse beads (disks) of diameter $d$ (equal to the lattice constant $a_0$) using a $p \approx 0.91$, the surface fraction of disks is about $\phi = 0.72$ ($\phi_{rcp} = 0.82$ for the random close packing in 2D).

Unlike previous simulations [20,24], we consider here the full elastic wave propagation (longitudinal and transverse modes) through a percolating network of $L \times L$ ($L = 70d$ in Fig. 5b) with the displacement field $\vec{r}_i(t)$ of a given bead at the site $i$ in the plane (x,y). To ensure the linear response to a transverse or shear displacement, the network is stretched by a static strain of $\varepsilon = 0.2$ via the four walls and the new positions of the beads define the initial stressed state $\vec{r}_i^{\,0}$. The displacement $\vec{r}_i(t)$ then satisfies the equation,

$$m(d^2\vec{r}_i/dt^2) = \sum_{i'} K_{ii'}(r_{ii'} - a_0)(\vec{r}_{ii'}/r_{ii'}) + \sum_{i'} \beta(d\vec{r}_{i'}/dt - d\vec{r}_i/dt) \qquad (1)$$

where $\vec{r}_i(t) = \vec{u}_i(t) + \vec{r}_i^{\,0}$ with $\vec{u}_i(t)$ the dynamic displacement ($u_i < 0.5 r_i^0$), $\vec{r}_{ii'}(t)$ is the distance vector between the bead *i* and its nearest neighbour *i'*, $a_0$ the spring length at rest (lattice constant), $K_{ii'}$ the fluctuating spring constants and $\beta$ a damping constant.

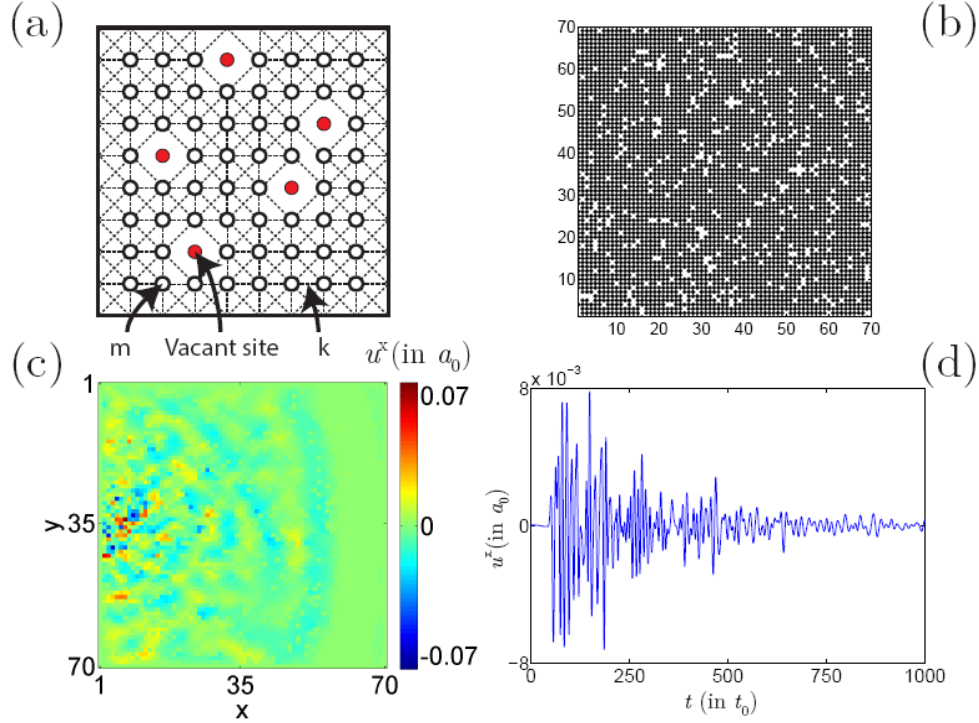

FIG. 5 (a) Sketch of the 2D percolating network of masses and springs with lattice constant $a_0$. (b) One realization of a 70x70 network with a number density $p \approx 0.91$ where the lattice sites are occupied by disks of diameter $d = a_0$, giving a the surface fraction of disks around 0.72. (c) Snapshot of a typical transmitted wave field $u^x(\vec{r},t)$ at $t = 46 t_0$ where the coherent wave (near $x \approx 60d$) is very weak due to scattering attenuation. (d) Temporal signal $u^x(\vec{r},t)$ recorded at position $x = 66d$, $y = 50d$, which lasts about $1000 t_0$ for a source duration of $5 t_0$.

Fig. 5c shows a typical wave field excited by a source consisting of three beads near the left wall ($x = 4d$), oscillating along the x-axis $\vec{u}_S^{\,x}(\vec{r}_0,t)$ with one sinus cycle at frequency $f = 0.2 f_0$ with $f_0 = (k_0/m)^{1/2}$ (one takes $m = 1$ and $k_0 = 1$). This snapshot of the dynamic displacement $u^x(\vec{r},t)$ at $t = 46 t_0$ ($t_0 = 1/f_0$) (with $\beta = 5.10^{-3}$ $m/t_0$) reveals a wave field composed of a weak coherent longitudinal wave (near $x = 60d$) followed by an irregular interference pattern likely due to multiply scattered waves; this observation is consistent with the temporal response $\vec{u}_i^{\,x}(t)$ detected at one site *i* (Fig. 5d). For the frequency used here, the longitudinal wavelength is $\lambda = V_P/f \sim 6d$ and attenuation length is $l_e \sim 10d$ dominated by scattering (data not shown), which indicates a weak multiple scattering regime ($l_e < 0.15L$).

To test the TR focusing, we take seventy beads near the right wall (at $x = 67d$) as receivers (i.e., TRM). We then time-reverse the displacement signals $\vec{u}_D(\vec{r},t)$ detected by each TRM element and back-propagate $\vec{u}_D(\vec{r},T-t)$ with $T$ the signal duration. Fig. 6a shows a snapshot of the $\vec{u}^x$ field near the focal time $t = T$. We observe that waves add up coherently near the source location but incoherently elsewhere, indicating that time-reversed waves do converge back to the source. We also check in Fig. 6b the recompressed displacement signal $\vec{u}_S^x(\vec{r}_0,T)$ at the source location; the sharp peak $\Psi(t)$ at $t \approx T$ confirms the TR focusing of elastic waves at the source.

*TR focusing in the nonlinear regime* –We now seek to model the irreversible sound-matter interaction effect, i.e., rearrangements of the contact network observed in TR experiments. This phenomenon is presumably related to the contact sliding between grains by the large source driving which modifies partly the initial contact network without visible motion of grains [18]. To simulate the modified contacts during the forward propagation step, we compare the maximum dynamic displacement $\vec{u}_i^{max}$ at each bead to its static stretching $\vec{r}_i^0$ and apply a local yield criterion $u_i^{max} \geq 0.02 r_i^0$. The larger the source amplitude, the larger the number of modified contact or the rearranged zone (data not shown). We ascribe to each modified contact a new value $k$ randomly chosen from the uniform distribution of stiffness and build up a new configuration of the network. Hence, we time-reverse the forward propagating signals through the *initial* network for a given source amplitude and back-propagate them in the accordingly *rearranged* network (note that in real experiments the initial network is modified during the forward propagation step).

Fig. 6c shows a snapshot of the displacement field $\vec{u}^x$ after the TR process at the focal time in the presence of rearrangements. Compared to the case without rearrangement at small source driving (Fig. 6a), we observe that the focusing spot around the source is less intense and that the wave field manifests some leakage of energy from the source location to elsewhere. Moreover, an amplitude decrease is found in the recompressed signal $\Psi^*(t)$ on the source at $t \approx T$ (inset of Fig. 6d). Fig. 6d shows the ratio $R$ of $\Psi^*_{max}$ to $\Psi_{max}$ obtained without rearrangement, calculated as a function of the source amplitude. The decrease of $R$ which may quantify the loss of fidelity in the TR process [6,24] supports well the experimental observations when the source driving is increased (Fig. 3c and Fig. 4).

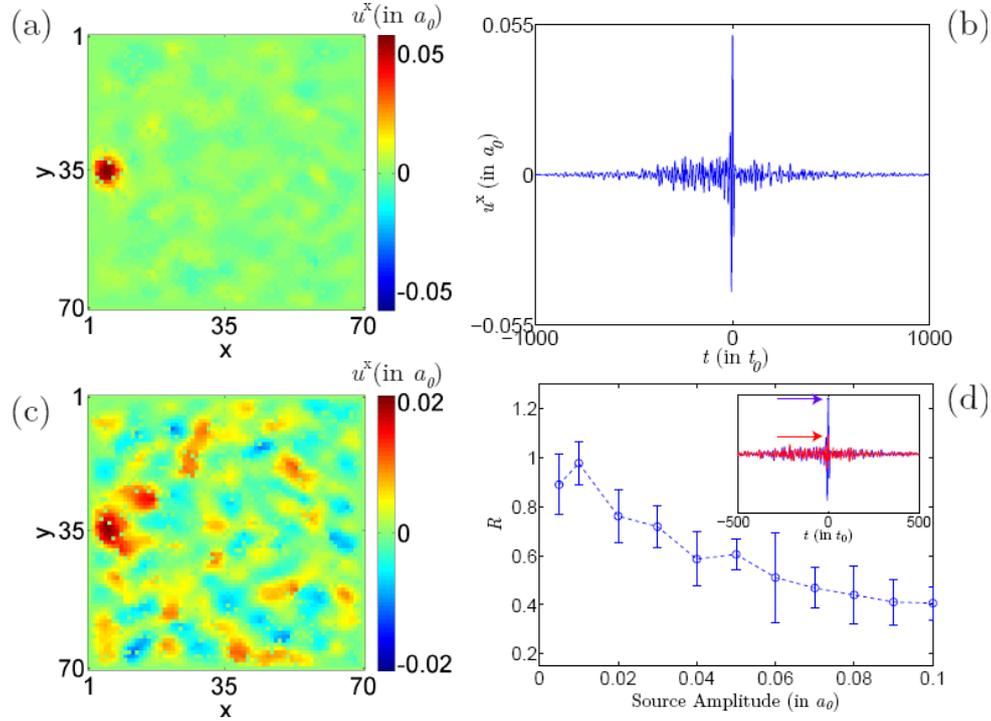

FIG. 6 Linear regime: (a) snapshot of a typical TR wave field $u^x(\vec{r},t)$ at the focal time $t = T$ where the waves add up coherently near the source location but incoherently elsewhere; (b) recompressed signal by TR at the source $\vec{r}_0$. Nonlinear regime: (c) snapshot of a TR wave field $u^x(\vec{r},t)$ at $t \approx T$ in the presence of rearrangements. TR focusing around the source location is less efficient than in the case without rearrangement shown in (a); (d) fidelity $R$ decreases as the driving amplitude increases. Inset: comparison of recompressed signals by TR focusing with (in red) and without (in blue) rearrangements.

## 4. Discussions and conclusion

As mentioned in the introduction, the acoustic TR invariance holds in a multiple scattering medium on the shortest wavelengths [1-7]. Consider a scalar displacement $u(\vec{r},t)$ that satisfies the dissipationless wave equation:

$$\rho(\vec{r})(\partial^2 u/\partial t^2) - \vec{\nabla} \cdot K(\vec{r})\vec{\nabla} u = 0 \qquad (2)$$

with $\rho(\vec{r})$ the density and $K(\vec{r})$ the elastic modulus of the heterogeneous medium. A source located at $\vec{r}_0$ transmits a short pulse $u_s(\vec{r}_0,t)$ (= $\delta(t)$, dirac-like) into the medium, the multiply scattered signals $u_d(\vec{r}_j,t) = h_j(t)$ are detected by receiver no. $j$ (TRMs) at $\vec{r}_j$ where $h_j(t)$ would denote the impulse response in linear acoustics. They are time-reversed to produce $u_d(\vec{r}_j, T-t)$ with *T* the signal duration and retransmitted into the medium. Taking into account the reciprocity principle, interchanging the source and the receiver does not alter the resulting wave field. The signal recreated at the source location can thus be written as $s(t) = \sum_j h_j(T-t) \otimes h_j(t)$, which is maximum at time *t = T* indicating a TR focusing in time at $\vec{r}_0$. In this study, we have shown by experiments (and simulations) in granular packings that the TR process using multiply scattered waves ($\lambda$ ~ 2-4*d*) applies correctly in the linear regime. These observations hence suggest that the elastic modulus *K* and elastic wave velocity *V* = (*K*/$\rho$)$^{1/2}$ defined in Eq. 2 within the continuum elasticity still holds on the local scale of a few grain size. This appears consistent with recent wave velocity measurements performed in stressed granular layers with thickness ~ 5*d* where the fluctuations in *V$_P$* remain less than 10% [26].

However, this TR invariance breaks down in the nonlinear regime with large source driving where the sound-matter interaction becomes irreversible (Fig. 3) due to the contact slipping associated with the nonaffine deformation in granular packings [14-16,18], particularly under shear (see Fig. 3c). Therefore, the propagation medium is different between the forward and backward steps, leading to a modified TR recompression signal $s'(t) = \sum_j h_j(T-t) \otimes g_j(t)$ at the source location where we denote by *g$_j$(t)* the impulse response of the rearranged granular network (the backward propagation is indeed performed in the linear regime). The decrease of TR focusing, or loss of fidelity, shown in Fig. 3c is related to the nonlinear sound-induced structure change, being very different from those observed in other TR experiments where the medium is changed (scatterer number, temperature) after the forward propagation [7]. Note nevertheless that TR invariance may still hold in a nonlinear regime as long as the medium is stationary (as in an ordinary elastic material) and before reaching the shock formation: if the whole harmonics generated during forward propagation are recorded, time-reversed and retransmitted, the initial wave can be reconstructed [27].

Time-reversal has also been used as a diagnostic tool to test and compare the sensitivity of particle motion and wave propagation to perturbations in the initial conditions or in the propagation medium [6,24]. In a multiple scattering system, particles and waves exhibit fundamentally different behaviours: particle motion is chaotic, incapable of returning to the source, while wave propagation is much more stable, despite TR invariance in both Newton's law and the wave equation (Eq. 2). The physical reason for this can be explained through the concept of ray splitting: a particle follows a well defined trajectory whereas waves travel along all ray directions (a huge number of trajectories) visiting scatterers in all positions within the scale of the wavelength. While a small perturbation can make the particle miss one obstacle or scatterer and completely change its future trajectory, the wave amplitude is much more stable due to coarse graining on the wavelength scale. *Granular acoustics* described here appears in between particle motion and wave propagation in terms of the sensitivity to perturbations, closely related to the discrete nature of the contact network with a finite coordination number $Z \sim 6$ in 3D bead packings (via Eq. 1) in which waves propagate along preferred paths. Further studies are planned to investigate the loss of fidelity as a function of the arrival time of multiply scattered waves [6].

Finally, we think that our study may be useful to those who are using time-reversal to seismic source locations or defect detections in fractured materials. In the context of seismic imaging, the application of TR requires that the difference between the real Earth (forward propagation) and the Earth model (backward propagation) can be neglected at the used wavelengths [8,9]. In laboratory experiments, an unknown source may be localized without the use of a numerical model when the reversed wave field is accessible by measurements. This is precisely the case for detecting fissures near the surface of a solid by TR process where harmonics generated by the nonlinear scatters, i.e., defects are selected and time-reversed to the source location [28]. However, our TR experiments in granular materials show that an important rearrangement of the medium by a large source driving may decrease the accuracy of the TR focusing and on the source location (Fig. 6) in fractured earth materials.

In conclusion, we believe that our time-reversal investigation in granular media may help to get a better understanding of the instability of wave scattering in nonlinear disordered media [29] and of the imprint of classical chaos on wave (quantum) systems [6,25]. It should also be useful for studying the source locations of seismic events and rupture in heterogeneous materials. We will further investigate TR focusing inside granular media, especially to learn to generate high-amplitude ultrasound at a particular position to trigger rearrangements.

**Acknowledgements**

We would like to dedicate this paper to the memory of Professor Roger Maynard from whom we received a lot of inspiration. He has always shown a great interest in both time reversal of waves in disordered media and acoustic propagation in granular matter. We also thank H. Sizun for the technical assistance to realize the TRM, R. Snieder, R. Pierrat and R. Carminati for helpful discussions. This work was supported by LABEX WIFI (Laboratory of Excellence ANR-10-LABX-24) within the French Program "Investments for the Future" under Reference ANR-10-IDEX-0001-02 PSL*. M.H. acknowledges the financial support from Université Paris Diderot, Y.Y. and X.J. from Université Paris-Est Marne-La-Vallée.